\documentclass[prb,preprint,showpacs]{revtex4}
\begin{document}

\title{Theoretical prediction and experimental study of a ferromagnetic shape memory alloy: Ga$_2$MnNi}
\author{S. R. Barman$^{1*}$, Aparna Chakrabarti$^2$,   Sanjay Singh$^1$, S. Banik$^1$,  S. Bhardwaj$^1$, P. L. Paulose$^3$, B. A. Chalke$^3$,  A. K. Panda$^4$, A. Mitra$^4$, A. M. Awasthi$^1$}
\affiliation{$^1$UGC-DAE Consortium for Scientific Research, Khandwa Road, Indore, 452017, India}
\affiliation{$^2$Raja Ramanna Centre for Advanced Technology, Indore, 452013, India} 
\affiliation{$^3$Tata Institute of Fundamental Research, Mumbai, 400005, India} 
\affiliation{$^4$National Metallurgical Laboratory, Jamshedpur,  831007, India} 

\begin{abstract}
We predict the existence of a new ferromagnetic shape memory alloy Ga$_2$MnNi using density functional theory. The martensitic start temperature ($T_M$) is found to be approximately proportional to the stabilization energy  of the martensitic phase ($\delta$$E_{tot}$) for different shape memory alloys. Experimental studies performed to verify the theoretical results show that  Ga$_2$MnNi is ferromagnetic at room temperature 
~and the $T_M$ and $T_C$ are 780\,K and 330\,K, respectively. Both from theory and experiment, the martensitic transition is found to be volume conserving that is indicative of shape memory behavior. 
\end{abstract}

\maketitle

\section{Introduction}
Ni$_2$MnGa exhibits both ferromagnetism and shape memory effect, and is a promising candidate for technological applications because of its high actuation frequency compared to conventional shape memory alloys.\cite{Murray00}  The unusually large strain caused by a moderate magnetic field (10\% at 1 Tesla)\cite{Murray00}  and the observation of giant magnetocaloric effect\cite{Marcos03} and large negative magnetoresistance\cite{Biswas05} in Ni$_2$MnGa have started intense research activity in ferromagnetic shape memory alloys (FSMA).  However, a major drawback  of Ni$_2$MnGa is its brittleness. So, the present challenge in FSMA research lies in the search for new materials that have magneto-mechanical properties superior to Ni$_2$MnGa, and preferably having high martensitic start temperature ($T_M$) and Curie temperature ($T_C$).  
~In recent past, different groups have attempted to find FSMA materials with properties superior to Ni$_2$MnGa. 
~Takeuchi {\it et al.} have studied a range of compositions in the Ni-Mn-Ga phase diagram and found that the martensitic transition temperature decreases as the magnetization increases.\cite{Takeuchi03} Although martensitic transition and inverse magnetocaloric effect have been reported recently in non-stoichiometric compositions of Ni-Mn-Sn, Ni-Mn-In, Ni-Co-Mn-In,\cite{Krenke05,Kainuma06,Sutou04}  these systems have not emerged as viable alternatives to Ni$_2$MnGa. 

Since related stoichiometric alloys like Ni$_2$MnAl, Ni$_2$MnIn, or Ni$_2$MnSn do not exhibit martensitic transition,\cite{Krenke05a} it is apparent that Ga plays an important role in making Ni$_2$MnGa a shape memory alloy. Substitution of Ga by In in Ni$_2$MnGa decreases $T_M$.\cite{Khan04} Thus, excess Ga may have a stabilizing effect on the martensitic phase. 
~Zayak {\it et al.} theoretically studied the role of Ga 4$p$ states in the stability of the martensitic phase of Ni$_2$MnGa.\cite{Zayak05} While considerable experimental work has been done on Ni and Mn excess Ni-Mn-Ga,\cite{Vasilev99,Khovaylo05,Liu05,Banik06}  this is not the case for Ga excess Ni-Mn-Ga.  Theoretical investigations so far have concentrated on the electronic structure in  Ni$_2$MnGa and related stoichiometric Heusler alloys and total energy calculations have been done to ascertain the stability of the martensitic phase.\cite{Ayuela99,Godlevsky01,Ayuela02,Bungaro03,Zayak03,Barman05}   

Here, using spin-polarized, full potential {\it ab initio} density functional theory, we establish a method to estimate the martensitic structural transition temperature and predict possible existence of a new FSMA  Ga$_2$MnNi.
~A tetragonal martensitic phase with $c/a$=\,0.83 is found to be lower in total energy ($E_{tot}$) compared to the cubic austenitic phase. 
~The  martensitic phase total energy is lower by 55 meV/atom (=$\delta$$E_{tot}$, {\it i.e.} the stabilization energy, which is the difference of $E_{tot}$ between the austenitic and martensitic phases). This value is larger than other FSMA materials studied by us.\cite{Barman05,Chakrabarti05,Barman07}  
~Based on our data and those in Refs.~\onlinecite{Ye97} and \onlinecite{Chen06}, we show that $\delta$$E_{tot}$ is approximately proportional to the martensitic transition temperature ($T_M$).
~$E_{tot}$ for the ferromagnetic state is lower than the paramagnetic state, showing  that Ga$_2$MnNi is  ferromagnetic. 
~Inspired by the theoretical prediction, Ga$_2$MnNi has been prepared, and it indeed exhibits a thermoelastic martensitic transition with  $T_M$=\,780~K, which is highest reported so far in the Ni-Mn-Ga family.  The Curie temperature ($T_C$) is \,330~K. X-ray diffraction (XRD) shows that the structure is monoclinic with $b$$\approx$\,7$\times$$a$, indicating the existence of modulation and hence, the possibility of magnetic field induced strain. 
Both from experiment and theory, we find that there is hardly any unit cell volume change across the martensitic transition, and this is strongly indicative of shape memory behavior.\cite{Bhattacharya} 

\section{Methodology}
The {\it ab-initio} relativistic spin-polarized full-potential-linearized-augmented-plane-wave (FPLAPW) method calculations  were performed using WIEN97 code\cite{Wien97} with the generalized gradient approximation for exchange correlation.  An energy cut-off for the plane wave expansion of 16 Ry is used ($R_{MT}$$K_{max}$=~9).  The cut-off for charge density is $G_{max}$=~14. The maximum $l$ ($l_{max}$) for the radial expansion is 10, and for the non-spherical part: $l_{max,ns}$=\,4.  The muffin-tin radii are Ni: 2.2488, Mn: 2.3999, and  Ga: 2.2488~a.u. The number of $k$ points for self-consistent field cycles in the irreducible Brillouin zone is 413 and 1063 in the austenitic and martensitic phase, respectively. 
~$E_{tot}$  consists of the total kinetic, potential and exchange correlation energies of a periodic solid.\cite{Weinert82}  The convergence criterion for the total energy $E_{tot}$ is  0.1 mRy, which implies that accuracy of $E_{tot}$  is $\pm$0.34 meV/atom. The  charge convergence is set to 0.001. The tetrahedron method for the $k$-space integration has been used. 

Polycrystalline ingots of Ga$_2$MnNi were prepared by melting appropriate quantities of the constituent metals of 99.99\% purity in an arc furnace under argon atmosphere and subsequently annealed in sealed quartz ampule wrapped in Mo foil at 873~K for 12 days, then at 723~K for 1~day and finally slowly cooled to room temperature. The differential scanning calorimetry (DSC) measurements were done by using TA Instruments MDSC model 2910 at a scan rate of 10$^{\circ}$/minute. Magnetization was performed using vibrating sample magnetometers (VSM) from Oxford Instruments and Lake Shore Cryotronics, Inc. Powder XRD patterns were obtained using Cu K$\alpha$ radiation with Rigaku XRD unit at a scan rate of 2$^{\circ}$/minute.
~Energy dispersive analysis of x-rays (EDAX) was done using  scanning electron microscope  with Oxford detector model  with 2\% accuracy by estimating the intensities of Ni, Mn and Ga K$\alpha$ characteristic lines (5.9  to 9.2~keV) 
~that are well separated and have small background by averaging over several measurements.

\section{Results and Discussion}

\subsection{Theoretical studies using FPLAPW method}

 
The positions of the atoms in the cubic austenitic phase of Ga$_2$MnNi are determined from the $E_{tot}$ calculations in the $L_{2_1}$ cubic structure that   consists of four inter penetrating f.c.c. lattices at (0.25,\,0.25,\,0.25), (0.75,\,0.75,\,0.75), (0.5,\,0.5,\,0.5) and (0,\,0,\,0) (Fig.~1a, b). The first two positions are equivalent ($8f$), whereas the other two are $4a$ and $4b$, respectively. In our notation, GaGaMnNi means that the two Ga atoms occupy (0.25,\,0.25,\,0.25) and (0.75,\,0.75,\,0.75) {\it i.e.} the $8f$ positions, while Mn and Ni are at (0.5,\,0.5,\,0.5) and  (0,\,0,\,0), respectively. Similarly, GaNiGaMn means that Ga atoms occupy inequivalent (0.25,\,0.25,\,0.25) and (0.5,\,0.5,\,0.5) positions, while Ni and Mn atoms are at (0.75,\,0.75,\,0.75) and (0,\,0,\,0), respectively. $E_{tot}$ has been calculated as a function of lattice constant ($a$) for all the different possible Ga positions (GaGaMnNi, GaGaNiMn, NiMnGaGa, NiGaMnGa, GaNiGaMn, GaMnGaNi), where the two Ga atoms occupy either symmetry equivalent or inequivalent points. $E_{tot}$ values for the inequivalent Ga structures (NiGaMnGa, GaNiGaMn and GaMnGaNi, Fig.~1b) are similar. The equivalent Ga structures (GaGaMnNi, GaGaNiMn and NiMnGaGa, Fig.~1a)  are also very close to each other in energy. The data have been fitted using a least square minimization routine using the Murnaghan equation of state (solid lines, Fig.~1c).
~The minimum $E_{tot}$ for the equivalent Ga  structures (arrow) is lower by 113 meV/atom compared to the inequivalent Ga structure (tick), unambiguously denoting  the former to be the stable structure of Ga$_2$MnNi in the austenitic phase.
~The  $E_{tot}$ minimum (arrow) is at $a$=\,11.285~a.u. (5.96~\AA) 
~ with the unit cell volume of 1437~a.u.$^3$ (Fig.~1c). 
~Furthermore, the formation energy of Ga$_2$MnNi is calculated by $E_{tot}$(Ga$_2$MnNi)-2$\times$$E_{tot}$(Ga)-$E_{tot}$(Mn)-$E_{tot}$(Ni). The formation energy 
turns out to be negative, comparable to Ni$_2$MnGa, indicating that the compound is stable. It should be noted that all the calculations shown in Fig.~1c have been performed in the ferromagnetic state, since this is the stable magnetic phase (discussed later).


The martensitic transition involves a structural transition from cubic to a lower symmetry phase with decreasing temperature.
In order to study this phase transition in Ga$_2$MnNi, our strategy is to calculate $E_{tot}$ as a function of a volume conserving tetragonal distortion  by varying $c/a$.  As $c/a$ is increased from the cubic value of unity, $E_{tot}$ increases (Fig.~1c). On the other hand, for $c/a$$<$1, $E_{tot}$ initially decreases and a minimum is obtained at $c/a$=\,0.83 
~(dashed tick). In the next step to reach the global $E_{tot}$ minimum in the martensitic phase, the unit cell volume is varied keeping $c/a$ fixed, 
~and the minimum is obtained at the unit cell volume of 1435.8 a.u.$^3$ with $a$=\,12.004, $c$=\,9.964~a.u. (Fig.~2a, dashed arrow). 
~Thus, although there is a large change in lattice constants (+6.4\% in $a$ and -11.7\% in $c$), there is almost no volume change between the austenitic  and the martensitic phases. 
 ~ $E_{tot}$ has been calculated for Ga$_2$MnNi in the  paramagnetic state in the martensitic phase using the optimized lattice constants. It turns out to be 156~meV/atom 
 ~higher than the ferromagnetic state. Thus, Ga$_2$MnNi has a ferromagnetic ground state.
 The total spin magnetic moment of Ga$_2$MnNi in austenitic (martensitic) phase is 3.04 (2.97) $\mu_B$. 
~The local moments of Mn, Ni, and Ga in the austenitic (martensitic) phase are 3.03 (2.87), 0.06 (0.16) and -0.05 (-0.05) $\mu_B$, respectively. On the basis of the condition that a volume conserving martensitic transition is the necessary and sufficient condition for shape memory behavior\cite{Bhattacharya} and that the ground state is ferromagnetic, 
~we predict Ga$_2$MnNi will behave as a ferromagnetic shape memory alloy.

The martensitic phase being the lower temperature phase, $E_{tot}$  for the martensitic phase is lower than the austenitic phase by 55 meV/atom. 
~Larger stabilization energy {\it i.e.} $\delta$$E_{tot}$ would imply greater stability of the martensitic phase and enhanced $T_M$. 
~From our earlier calculations, $\delta$$E_{tot}$ (experimental $T_M$) is found to be 3.6 (210~K), 6.8 (270~K) and  39 (434~K) meV/atom for Ni$_2$MnGa, Mn$_2$NiGa and Ni$_{2.25}$Mn$_{0.75}$Ga, respectively.\cite{Barman05,Chakrabarti05,Barman07,Banik07}   Here, we report similar calculations for Ni$_2$MnIn and Ni$_2$MnAl. The optimized lattice constants of Ni$_2$MnAl and Ni$_2$MnIn  (5.79 and 6.06\AA,~ respectively) are in good agreement with experiment: 5.83 and 6.08\AA.\cite{Kudryavtsev04}    Although their off-stoichiometric compositions exhibit martensitic transition, it is well known that neither of these Heusler alloys undergo martensitic transition.\cite{Krenke05a} Interestingly for Ni$_2$MnIn, $\delta$$E_{tot}$ turns out to be almost zero (0.34 meV/atom) within the theoretical accuracy limit, while $\delta$$E_{tot}$ for Ni$_2$MnAl is negative (-0.94~meV/atom). These values of $\delta$$E_{tot}$ indicate that the  martensitic phase in Ni$_2$MnAl and Ni$_2$MnIn is not stable and so  martensitic transition will not occur. This is in agreement with experimental data and earlier theoretical work.\cite{Ayuela99,Godlevsky01}

From the above data, a correlation emerges between $\delta$$E_{tot}$ and  $T_M$. Conceptually, this is understandable since larger $\delta$$E_{tot}$ implies higher stability of the  martensitic phase at zero temperature. A first order transition to the austenitic phase would occur when with increasing temperature, the martensitic phase energy (defined by the energy minimum in Fig.~2a) would increase to reach the energy minimum for the austenitic phase. This means with increasing temperature, to undergo the martensitic transition, the energy of the martensitic phase has to overcome $\delta$$E_{tot}$ and this would be directly related to $k_B$$T_M$.   
~A similar concept has been used in Ref.~\onlinecite{Chakrabarti05},  where taking  $\delta$$E_{tot}$$\propto$$k_B$$T_M$, the increase in $T_M$ between Ni$_2$MnGa and Ni$_{2.25}$MnGa could be explained. This expression should be generally valid, and this indeed seems so for TiNi (45, 333), TiPd (95, 783) and TiPt (155, 1343).\cite{Ye97} The numbers in bracket indicate $\delta$$E_{tot}$ and $T_M$ in meV/atom and K, respectively, as taken from Ref.~\onlinecite{Ye97}. Similar trend is obtained for Ni excess Ni-Mn-Ga.\cite{Chakrabarti05,Chen06,Banik07}

In Fig.~2b, $T_M$ versus $\delta$$E_{tot}$ for all the shape memory alloys discussed above are plotted; $T_M$ is taken to be zero for Ni$_2$MnIn and Ni$_2$MnAl. It is highly significant that although the theoretical data are from three different groups\cite{Barman05,Chakrabarti05,Barman07,Ye97,Chen06} on two different types of shape memory alloys and the methods of calculation are different, an approximately linear relation between $T_M$ and $\delta$$E_{tot}$ is evident. Thus the  validity of the expression $\delta$$E_{tot}$$\propto$$k_B$$T_M$ is established. A rather good straight line fit through the data for TiX (=\,Ni, Pd, Pt)\cite{Ye97} is obtained (Fig.~2b). Since the Ni-Mn-X (X=\, Ga, In, Al) FSMA's are different from TiX, a separate straight line is fitted. The quality of the fit is similar to TiX; except for data around 200~K. This is possibly because of the existence of modulated structures this $T_M$ range, which is not considered in theory. 
~From the fitted line, $T_M$ for Ga$_2$MnNi is estimated to be about 570~K (filled circle), corresponding to its $\delta$$E_{tot}$=\,55~meV  (Fig.~2b). 

It is generally believed that $T_M$ would increase with the valence electron per atom ratio ($e/a$). However, this relation is of limited applicability and breaks down in many cases: for example,  Ni$_2$MnGa, Ni$_2$MnIn and Ni$_2$MnAl all have the same $e/a$ (=\,7.75), but only Ni$_2$MnGa exhibits a martensitic transition. TiX (X=\,Ni, Pt, Pd) has the same $e/a$ (=\,6.5), but their $T_M$ is very different. In Ni-Mn-Ga-In, although $e/a$ is same, $T_M$ changes.\cite{Khan04} For Ni$_{2-x}$Mn$_{1+x}$Ga between $x$=\,0.25 to 1, we find that as $e/a$ decreases from 7.31 to 6.75, $T_M$ increases from 37 to 270~K.\cite{Liu05,Banik08a} 
~For the alloys shown in Fig.~2b, the absence of any correlation between $T_M$ and $e/a$ is shown as an inset. In contrast,  the present approach explains all the above observations. For example, $\delta$$E_{tot}$ decreases from 3.6 meV/atom to zero between Ni$_2$MnGa and Ni$_2$MnIn, which explains the decrease in $T_M$ with In doping and the absence of a martensitic transition in Ni$_2$MnIn. Higher $\delta$$E_{tot}$ in Mn$_2$NiGa rationalizes why its $T_M$ is higher than Ni$_2$MnGa, although its $e/a$ (=\,6.75) is lower. Thus, the proportionality of $T_M$ with $\delta$$E_{tot}$ is of more general validity, since it has a theoretical foundation that involves all electron {\it ab-initio} calculations, unlike the phenomenological relation between $T_M$ and $e/a$. In fact, this approach to determine the transition temperature should be applicable to any first order structural transition. 

\subsection{Experimental studies}


~Differential scanning calorimetry on polycrystalline ingots of Ga$_2$MnNi shows a clear signature of a first order martensitic transition with $T_M$=\,780\,K and austenitic start temperature ($A_s$)  of 790\,K (Fig.~3a). 
~The experimental $T_M$ is considerably higher than the theoretically predicted value, and a possible reason is discussed below. The latent heat of the transition turns out to be  about 2.35\,KJoule/mole, which is similar to that reported for Ni excess Ni-Mn-Ga, for example, Ni$_{2.24}$Mn$_{0.75}$Ga.\cite{Banik07} The difference in the width of the heating and cooling thermograms could be related to the kinetics of the structural transition. 
 EDAX measurements from   different regions of 30$\mu$$\times$30$\mu$ area  as well as the back scattered image show that the specimen is homogeneous. The average composition turns out to be Ga$_{1.9}$Mn$_{1.08}$Ni$_{1.02}$. 
~ In agreement with theory, the isothermal $M-H$ curve at 2.5~K  shows that Ga$_2$MnNi is indeed ferromagnetic (Fig.~3b). The hysteresis loop is not clearly observed because the coercive field is small ($\approx$25~mT). Such small coercive fields have been reported for other Ni-Mn-Ga alloys.\cite{Tickle99,Banik06} The saturation field is  1~T and the saturation moment is 1~$\mu_B$/f.u.  $M(T)$ at low field gives $T_C$=\,330~K (arrow, Fig.~3c). This implies that the martensitic transition occurs in the paramagnetic state  and expectedly $M(T)$ shows no change across $T_M$.
If should be noted that the saturation moment of 1\,$\mu_B$/f.u. is less than the theoretically calculated moment of about 3\,$\mu_B$/f.u.  The reasons for this disagreement could be that the actual sample has Mn excess, which might  cause Mn clustering leading to antiferromagnetic coupling between Mn atom pairs, as has been observed for other Mn excess systems.\cite{Barman07,Enkovaara03,Banik08} Moreover, note that the theory does not consider the actual monoclinic structure (discussed below) which might favor a different magnetic ground state with anti-parallel coupling between Mn atoms. 

The x-ray diffraction (XRD) pattern corresponding to the  austenitic phase 
~has been simulated by the Le Bail fitting procedure, and the structure is clearly cubic $L_{2_1}$. 
~The relative intensity of the (200)  peak compared to the (111) peak (shown in an expanded scale in Fig.~3d) confirms that the Ga atoms occupy the equivalent $8f$ position, in agreement with theory (Fig.~1). The experimental lattice constant ($a_{aus}$=\,5.84~\AA)~ is close to the calculated value (5.96~\AA). However, the martensitic phase XRD pattern is more complicated than tetragonal and can be indexed by a monoclinic phase ($P2/m$ space group) with $a$=\,4.31, $b$=\,29.51 and $c$=\,5.55\,\AA,~ and $\beta$=\,90.49. Since $b$$\approx$\,7$\times$$a$, a seven layer modulation may be expected, and such structures with monoclinic or orthorhombic symmetry that exhibit modulation has been reported for  Ni-Mn-Ga.\cite{Brown06} Magnetic field induced strain has been observed in Ni-Mn-Ga for structures that exhibit modulation.\cite{Murray00} The $c/a$ for this monoclinic cell (that can be compared to the theoretical $c/a$= 0.83 for the tetragonal structure) is obtained by $c/a$= 5.55/(4.31$\times$$\sqrt{2}$)= 0.91. Thus, the agreement between experimental and theoretical $c/a$ is reasonable, considering that a simplified structure is used in theory.  

However, the most important point is that the experimental unit cell volume of the martensitic phase is within 1\%  of that of a comparable austenitic cell given by 7$\times$$a_{aus}^3$/2. This shows that the unit cell volume hardly changes between the two phases, which is a necessary condition for a shape memory alloy. Thus, a unit cell volume conserving martensitic transition with small width of hysteresis (Fig.~3a) and presence of modulation  indicate that Ga$_2$MnNi is indeed a FSMA material. 


\section{Conclusion} 
The modulated martensitic structure of Ni-Mn-Ga is complicated and a controversy exists even about the structure of the  well studied Ni$_2$MnGa.\cite{Brown06}  Atomic positions have not yet been determined for the  monoclinic structure. 
~Under such circumstances, our work is  important because it shows that a new FSMA material can be predicted by  computing the energy cost of formation of the martensitic phase in a simpler tetragonal structure. 
The present work demonstrates that a new FSMA material can be predicted by determining the energy stability of a tetragonal martensitic phase with respect to the cubic austenitic phase. 
~This approach is successful because, although the modulated phase involves a large unit cell, the atoms are generally displaced only by a small amount from their positions compared to the tetragonal structure.\cite{Brown06}
~Since the tetragonal structure is not computationally demanding, precise  calculations can be performed for lattice constant optimization in the lowest energy magnetic state.\cite{Barman05,Chakrabarti05,Barman07} Thus, the total energy difference can be determined with sufficient accuracy and thus $T_M$ estimated. However, difference in $T_M$ between experiment and theory could occur, as in this case, possibly because the latter does not consider the actual structure. In this context, it is to be noted  (Fig.~2) that a subtle change in $\delta$$E_{tot}$ can substantially alter the $T_M$ value.   
~Theory thus provides an important starting point for the experimentalists, and experimental inputs can be used to further refine the theory. 
~
A direct proof of the FSMA behavior is the movement of twins with magnetic field and the actuation behavior. So, further work on the magneto-mechanical  behavior of Ga$_2$MnNi is in progress. 
~Prediction of new materials in the quest for better properties is the need of the hour in FSMA research 
~and the present work aims towards that. 

\section {Acknowledgment} 
P. Chaddah, V. C. Sahni, A. Gupta, S. M. Oak, and K. Horn are thanked for encouragement. Help from the scientific computing group of the Computer Centre, RRCAT is acknowledged. I. Bhaumik, P. K. Mukhopadhyay and  R. J. Chaudhary are thanked for useful discussions. Funding from Ramanna Research Grant, D.S.T. and Max-Planck Partner Group Project is acknowledged.

\noindent $^*$E-mail: barman@csr.ernet.in




\end{document}